\documentclass[journal=langd5,manuscript=article]{achemso}
\usepackage[sort&compress,numbers,super]{natbib}
\usepackage{caption}
\usepackage{float}
\usepackage{geometry}
\usepackage{setspace}
\usepackage{xkeyval}
\usepackage{achemso}
\usepackage[version=3]{mhchem}
\usepackage{mathptmx}
\usepackage{graphicx} 
\usepackage{bm}
\usepackage{hyperref}
\usepackage[english]{babel}

\author{Jasper N. Immink}
\phone{+46-46-2223677}
\email{jasper.immink@fkem1.lu.se}
\affiliation{Division of Physical Chemistry, Lund University, Lund, Sweden.}
\author{J. J. Erik Maris}
\affiliation{Inorganic Chemistry and Catalysis Group, Utrecht University, Utrecht, the Netherlands.}
\author{Peter Schurtenberger}
\affiliation{Division of Physical Chemistry, Lund University, Lund, Sweden.}
\author{Joakim Stenhammar}
\affiliation{Division of Physical Chemistry, Lund University, Lund, Sweden.}

\title{Using patchy particles to prevent local rearrangements in models of non-equilibrium colloidal gels}

\begin{document}

\begin{abstract}
\noindent Simple models based on isotropic interparticle attractions often fail to capture experimentally observed structures of colloidal gels formed through spinodal decomposition and subsequent arrest: the resulting gels are typically denser and less branched than their experimental counterparts. Here we simulate gels formed from soft particles with directional attractions (``patchy particles''), designed to inhibit lateral particle rearrangement after aggregation. We directly compare simulated structures with experimental colloidal gels made using soft attractive microgel particles, by employing a ``skeletonization'' method that reconstructs the 3-dimensional backbone from experiment or simulation. We show that including directional attractions with sufficient valency leads to strongly branched structures compared to isotropic models. Furthermore, combining isotropic and directional attractions provides additional control over aggregation kinetics and gel structure. Our results show that the inhibition of lateral particle rearrangements strongly affects the gel topology, and is an important effect to consider in computational models of colloidal gels.
\end{abstract}

\section{Introduction}
\noindent The aggregation of suspended particulate matter due to attractive interactions can cause the formation of a volume-spanning network,~\cite{Jullien1994} which has been observed in various materials such as foodstuffs,~\cite{Dickinson2005} proteins,~\cite{Durand2010} and cement.\cite{DelGado2016} The process of gelation relies on spinodal decomposition into particle-poor and particle-rich phases, where the latter arrests to form a percolating network whose precise nature depends on parameters such as the particle volume fraction $\phi$, the attraction strength, and the quench speed.\cite{Dhont1995,Weitz2005,Schurtenberger2007,Wiltzius1986,Russel1993,Malins2012} This characteristic spinodal decomposition and subsequent arrest behavior can be recreated with computational models involving strong, isotropic attractions between spherical particles.\cite{Bartlett2011} However, using such models often leads to discrepancies compared to the experimentally observed gel structures. For example, when using isotropic models, the formation of percolated network structures has been found to require $\phi > 0.07$,~\cite{Dickinson2013} while experimentally, percolation can be found for volume fractions as low as $\phi = 0.025$.~\cite{Opheusden1995,Weitz2002,Tanaka2015,Immink2019} Several computational studies have considered the absence of hydrodynamic interactions (HIs) as an explanation for the observed discrepancies between ``dry'' models and experiments.~\cite{Araki2000,Tanaka2010,Luijten2011,Morris2012,Tanaka2015,Swan2015} While HIs have been shown to slow down the compactification of clusters, leading to more elongated local clusters in gels and percolation at $\phi = 0.04$,\cite{Tanaka2015} a recent analysis of simulations with and without HIs by de Graaf \emph{et al.}~\cite{Hermes2019} indicates that the corresponding gel \textit{structures} are in fact very similar as long as the observation time is rescaled to account for the significantly different \textit{dynamics} induced by HIs. Thus, it appears that structural discrepancies between experiments and modelling are not fully accounted for by HIs, and that other effects need to be incorporated into the models. 

One possible explanation for more branched network structures seen in experiments is the inhibition of lateral particle rearrangements after initial aggregation. Particle surface roughness can lead to interlocking and an effective rotational friction between two particles,\cite{Isa2018} hampering lateral particle movement. Such rearrangements are not penalized when using interaction potentials that only depend on the separation between particles, leading to a stronger coarsening and densification of the particle network after aggregation.~\cite{Morris2012,Tanaka2015,Swan2019} Furthermore, for soft particles composed of crosslinked polymer networks in water, so-called \emph{microgels},~\cite{Pelton1986,Harden2010} polymer ends on the particle surfaces can exhibit solvent-avoiding behavior at elevated temperatures.~\cite{Jiang1996} The nature of the interparticle attraction between microgel particles is complex, and contains contributions from van der Waals attraction\cite{Vincent2004} and the interaction between solvent-avoiding polymers on particle surfaces,\cite{Bureau2010} where the latter can furthermore cause polymer entanglement upon colloidal aggregation. This “locking in” of the local configuration upon aggregation is not effectively modeled using isotropic potentials. In this Article, we propose a model of such soft microgel particles based on an overlay between an isotropic repulsion and a directional attraction, so-called ``patchy particles'', as a tool to hinder lateral particle reorganization after gelation. This class of models has been extensively studied in the context of \emph{equilibrium} systems such as polymer gels, associating fluids, proteins, and patchy colloids,~\cite{Kob2010, Sciortino2010,Likos2011,Sciortino2006,Zdeborova2008,Likos2015,Zaccarelli2017,Kahl2012,Truskett2019} in addition to one very recent study of non-equilibrium colloidal gels formed \textit{via} depletion attraction.~\cite{Swan2019} We show here that such patchy particle models are capable of capturing the experimentally observed structures in non-equlibrium colloidal gels formed through arrested spinodal decomposition of soft microgel particles interacting \emph{via} short-range attractive interactions and soft repulsion.~\cite{Immink2019} By using a novel analysis technique based on a ``skeletonization'' of the gel backbone,~\cite{Chu1994} we compare structural properties between gels formed in simulations with those obtained from experiments on soft poly(N-isopropylacrylamide) (pNIPAm) particles that aggregate following an initial temperature-induced collapse. We show that using a directional model with a sufficient number of attractive patches promotes the formation of highly branched gel structures which reflect the experimentally obtained structures even in the absence of hydrodynamic interactions. Our results suggest that inhibited lateral particle rearrangements due to polymer entanglement and surface roughness play an important role in determining the structural signatures of colloidal gels made from soft particles, and that models based on directional attractions are suitable for describing such systems. 

\section{Results and discussion}
\begin{figure}
	\centering
	\includegraphics{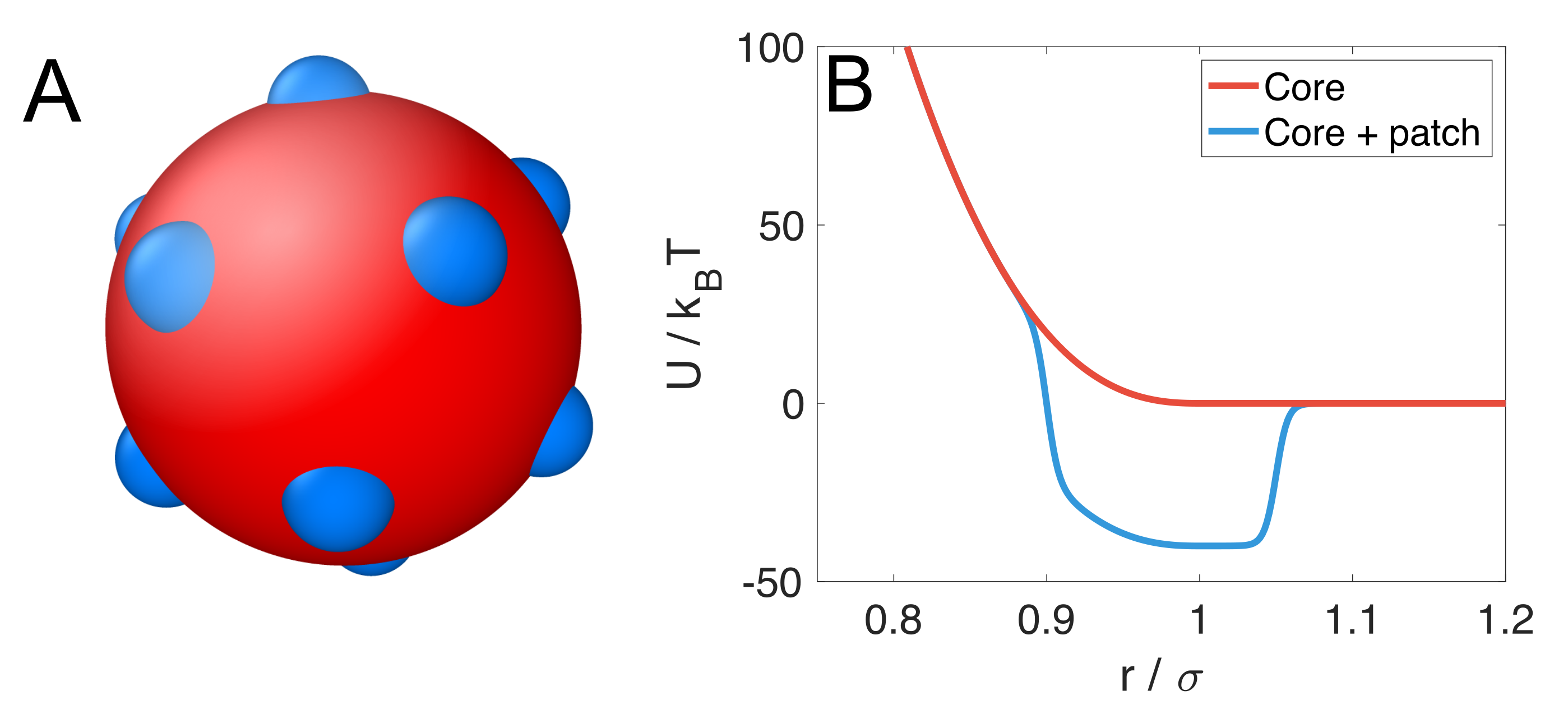}
	\caption{(A) Patchy particle with $N_p = 12$ patches arranged in an icosahedral geometry. (B) The repulsive Hertzian interaction potential as a function of the center-to-center separation $r$ (red), and the corresponding total (repulsive + attractive) potential, with the latter representing the case of two patches perfectly facing each other.}
	\label{Fig1}
\end{figure}

As described in further detail in the Methods section, the soft repulsion between particles is modelled as an isotropic Hertzian potential with an interaction diameter $\sigma$, which we use as our basic unit of length; this model has been used extensively to model soft microgel particles.~\cite{Mohwald2010,Schurtenberger2013} The repulsive particles are decorated with $N_p$ attractive patches ($4 \leq N_p \leq 12$) of strength $40 k_B T$, with $k_B T$ the thermal energy, strong enough to ensure irreversible aggregation. Each patch is described through a rounded square-well-like potential, centered around regularly spaced points on the particle surface (see Fig.~\ref{Fig1}). In addition, we studied the corresponding \textit{isotropic} attraction model where the same attractive potential instead acts along the entire particle surface, leading to each bond having the same strength as for the patchy model but with no directionality. The model was solved using Brownian dynamics (BD) simulations of $N = 10^4$ particles at a volume fraction of $\phi = 0.05$, matching the approximate experimental density; for further details, see the Methods section. Note that the directionality imposed on the bonds in our model is a way to prevent lateral particle rearrangements after aggregation: the underlying experimental interaction potential is still effectively isotropic. Thus, in order to accurately model the experimental system it is crucial that $N_p$ is large enough to ensure that virtually all particle collisions lead to a bond (\emph{i.e.}, that aggregation remains diffusion limited), while still being small enough to prevent lateral rearrangements through the formation of multiple bonds per particle pair. Alternative ways of constraining the particle bonds after aggregation proposed previously is the use of three-body potentials~\cite{Srikanth2009,Kob2010,DelGado2014} or more complex simulation algorithms that create surface-localized, irreversible bonds after collision.~\cite{Dickinson1996,vanVliet2001} While aiming towards the same goal, these approaches are computationally more complex than models based on attractive patches. 

The simulation results were compared to experiments on gels formed from fluorescently labeled pNIPAm particles at $\phi = 0.05$ with approximately 10 minutes aging time. The gel backbone was reconstructed from Confocal Laser Scanning Microscopy (CLSM) images using a ``skeletonization'' algorithm.~\cite{Chu1994, Fratzl2013} The experimental backbone structures were then compared with those obtained from BD simulations after the same effective aging time of 10 minutes. Further details of the experimental and analysis procedures are given in the Methods section. \\
\begin{figure}
	\centering
	\includegraphics[width=\textwidth]{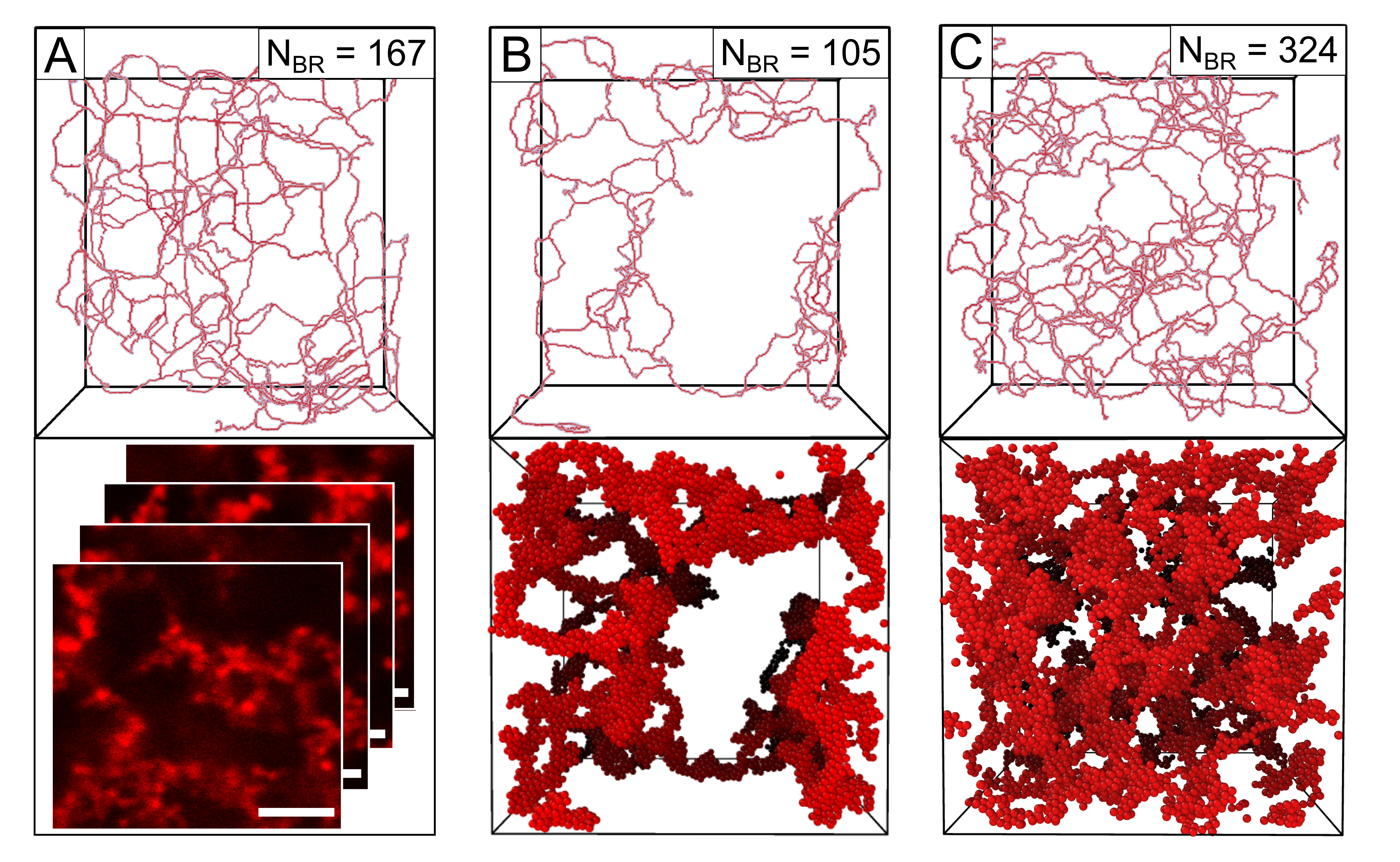}
	\caption{Gel backbones reconstructed using the skeletonization algorithm applied to (A) a stack of averaged CLSM images and (B,C) simulations using either (B) an isotropic interaction potential or (C) a directional potential with $N_p = 12$. The number of branch points in the backbone $N_{BR}$ is shown in the insets, where the system volume is $(47.1\sigma)^3$ in both experiments and simulations. The bottom row shows the corresponding particle configurations, with the CLSM data showing a number of 2D images from the 3D stack. The scale bar is 12~$\sigma$ long.}
	\label{Fig2}
\end{figure}
\indent Fig.~\ref{Fig2} shows the gel backbone structures retrieved from both experimental gels and from BD simulations of the isotropic and patchy particle models. Here, $N_{BR}$ is the number of branch points in the backbone, averaged over four simulations or 200 experimental image stacks. Comparing the latter two images (Figs.~\ref{Fig2}B and C), it is clear that the directional interaction yields a significantly more branched and less open structure than the isotropic interaction, due to the inhibition of lateral diffusion after initial aggregation, and can better approach the experimental results. In order to further investigate the importance of directional bonds on the final gel structure, we performed simulations using four different valencies: $N_p = 4,6,8$ and $12$, arranged in tetrahedral, octahedral, cubic and icosahedral geometries on the particle surface. In Fig.~\ref{Fig3}A, we follow the time evolution of the gelation process by calculating the average number of nearest neighbors $\langle Z \rangle$ within a distance $\sigma$ of each particle. It is clear that the introduction of patches strongly decreases the long-time value of $\langle Z \rangle$ by inhibiting particle reorganization after initial aggregation, an effect not due to the finite number of binding sites alone: even with $N_p = 12$, $\langle Z \rangle$ saturates at a value of $\approx 4.5$, which is significantly lower than the corresponding value of  $\langle Z \rangle \approx 7.5$ observed for the isotropic model. The initial aggregation kinetics are somewhat slowed down for the patchy particles compared to the isotropic ones, indicating that the directionality of the interactions alters the diffusion-limited dynamics of the isotropic system, since the patchy particles need to align their binding sites before aggregating; we will see below how this unwanted effect can be overcome through the overlay between isotropic and directional interactions. The structure factors $S(q)$ of the simulated gel structures are shown in Fig.~\ref{Fig3}B. In accordance with our previous observations, $S(q)$ for the isotropic system shows significantly higher values than the corresponding patchy particle gels at intermediate $q$, reflecting the thicker strands in the former. At high $q$, the specific patch geometries induce local order that causes peaks reflecting the various local structures. Finally, the low-$q$ values of $S(q)$ again clearly show the smaller mesh size of the gels formed from directionally interacting particles as compared to isotropic ones. 

In Fig.~\ref{Fig4}A, we analyze the tortuosity of the gel skeleton, $\xi$, defined as
\begin{equation}\label{eq:xi_def}
\xi = \left\langle\frac{\lambda(A,B)}{\lambda_{Euc}(A,B)}\right\rangle
\end{equation}
where $A$ and $B$ are start and end points of paths along the gel backbone intersecting respectively the upper or the lower box face in any Cartesian direction, $\lambda(A,B)$ is the length of this path, while $\lambda_{Euc}(A,B)$ is the Euclidian distance between $A$ and $B$; the angular brackets denote averaging over all such paths for a given configuration. This measure is often used to characterize flow rates through porous catalysts~\cite{Hu2017} or blood vessel networks,~\cite{Banaee2016} where a large density of branch points, high degree of branching and non-erratic strands decrease the tortuosity of the network.~\cite{Gladden1992,Liu2006,Harvey2018} In Fig.~\ref{Fig4}A we observe a clear decrease in tortuosity as the gel coarsens, due to a gradual straightening of gel strands. At long times, $\xi$ relaxes to a value that is lower for higher $N_p$, as a consequence of the higher node density and degree of branching. The experimentally obtained value after 10 minutes of aging (dashed line in Fig.~\ref{Fig4}A) is furthermore very close to the asymptotic value from the simulations with $N_p = 12$, indicating a macrostructural similarity in strand geometry between experiment and simulation.
\begin{figure}
	\centering
	\includegraphics[width=\textwidth]{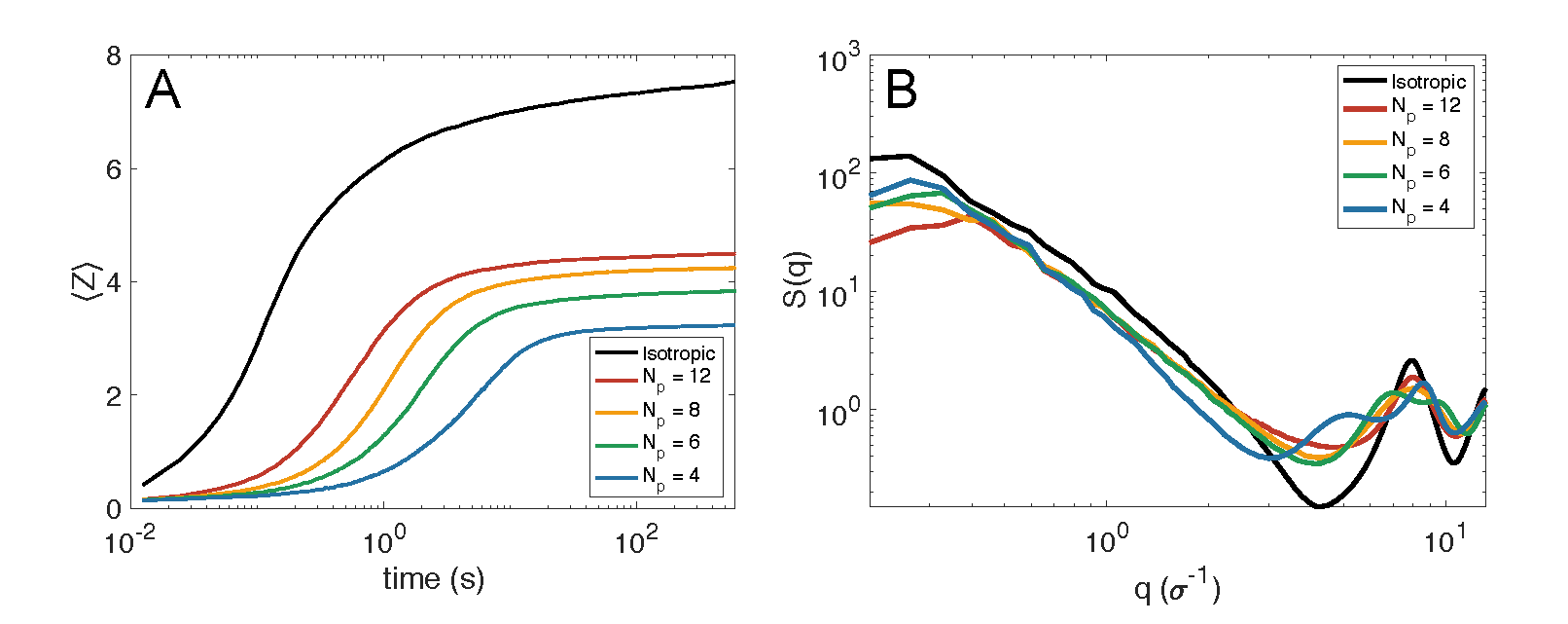}
	\caption{Time evolution of (A) the average number of nearest neighbors per particle $\langle Z \rangle$, (B) the structure factors $S(q)$ of the final configuration of the gel. }
	\label{Fig3}
\end{figure}
\begin{figure}
	\centering
	\includegraphics[width=\textwidth]{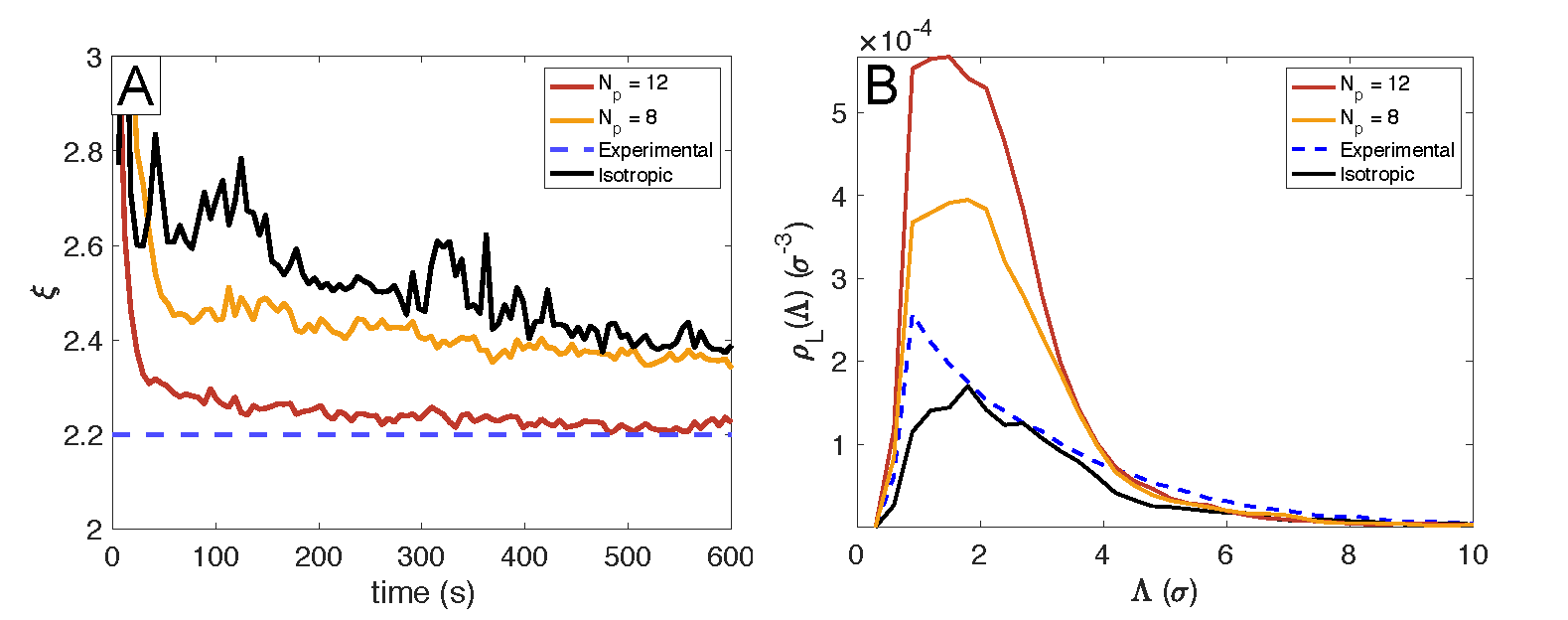}
	\caption{(A) The tortuosity $\xi$, defined in Eq.~\eqref{eq:xi_def}, as a function of time for the patchy particle simulations with $N_p = 12$ and $N_p = 8$, and isotropic particle simulations. The experimentally obtained value, as obtained from approximately 200 experimental image stacks at 10 minutes aging time, is indicated by the dashed line. (B) Link density $\rho_L(\Lambda)$ as a function of the link length $\Lambda$ for the final gel structures formed using the isotropic and the directional potential with $N_p = 12$ and $N_p = 8$, as obtained from the final 150 frames (approx. 30 seconds) averaged over four separate simulations. The experimental value (dashed line) was obtained as in (A). $N_{BR}$ denotes the average total number of branch points in the system. Data for low-valency patchy particle simulations are omitted due to high statistical noise.}
	\label{Fig4}
\end{figure}
We further quantify the gel structure by viewing it as a set of \emph{nodes} connected by \emph{links}, with $\rho_L(\Lambda)$ the density of links of length $\Lambda$, as shown in Fig.~\ref{Fig4}B. The patchy potential clearly leads to a significant increase in the density of short links, corresponding to a more strongly branched network, an effect which becomes stronger with increasing $N_p$. The experimental $\rho_L(\Lambda)$ curve and the total number of branch points $N_{BR}$ is furthermore significantly higher than its isotropic model counterpart, reflecting the sparse network structure seen in Fig.~\ref{Fig2}B. The slight shift in peak position and width in simulations compared to the experimental curves is likely caused by optical limitations inherent to CLSM.~\cite{Egelhaaf2008} 

\begin{figure}
	\centering
	\includegraphics{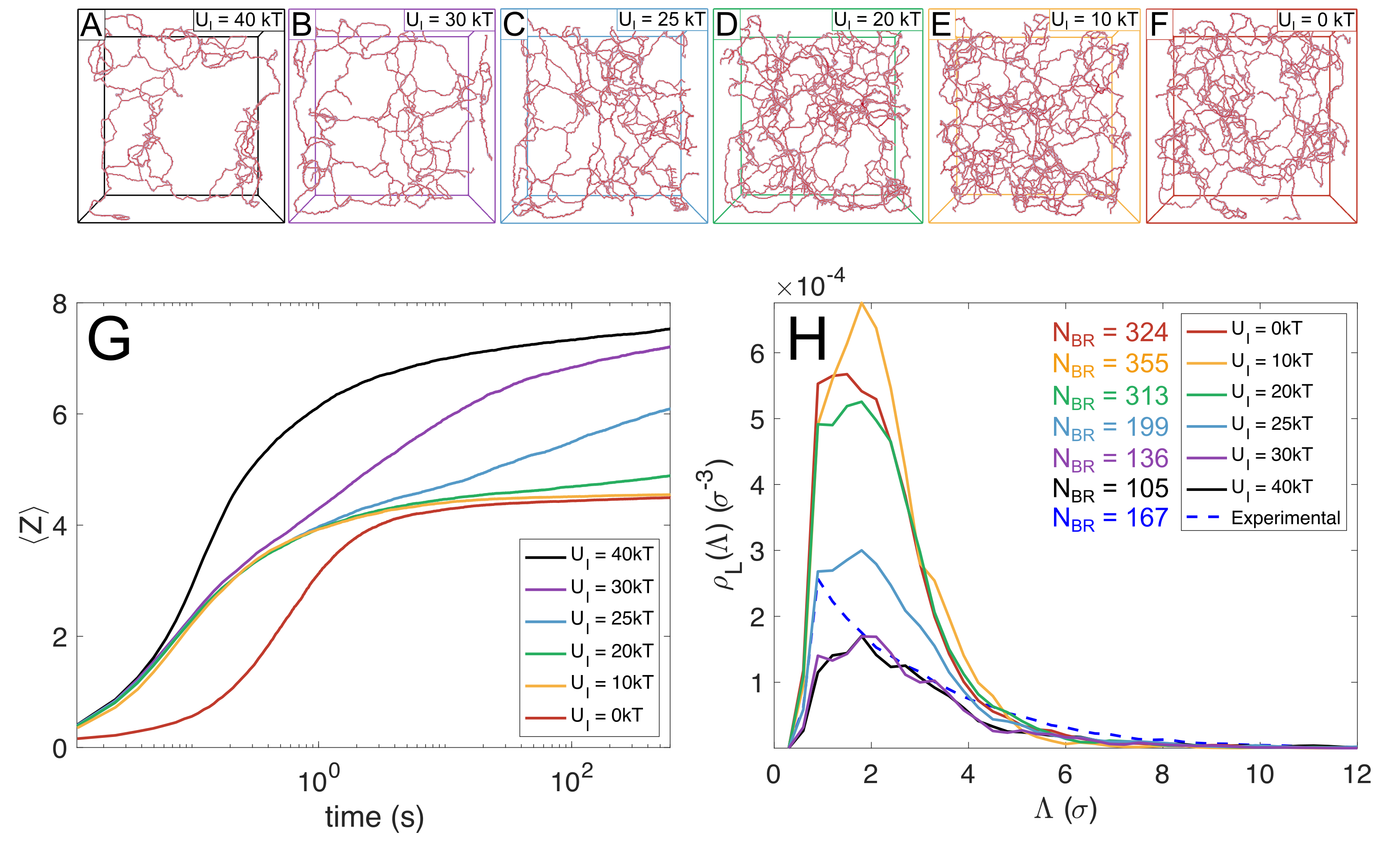}
	\caption{Overlaying patchy and isotropic potentials allows for increased control of structure and aggregation kinetics. (A)--(F)~Simulated backbone structures after 600 seconds aging time. (G)~Time evolution of $\langle Z\rangle$ for the systems in (A)--(F). (H) The corresponding link density $\rho_L(\Lambda)$; the dashed line shows the same experimental curve as in Fig.~\ref{Fig4}B. In all cases, the number of patches is $N_p = 12$, and the sum of the patchy and isotropic potential depths is kept constant at $U_P + U_I = 40 k_B T$. }
	\label{Fig5}
\end{figure}

As previously observed in Fig.~\ref{Fig3}A, a limitation of using a directional interaction model to describe intrinsically isotropic particles is the possibility of collisions not leading to bond formation, which stops the initial aggregation kinetics from being accurately described. In Fig.~\ref{Fig5}, we therefore depict the results from an overlay of the patchy attraction (of strength $U_P$) with an isotropic attraction (of strength $U_I$), constraining the total energy per bond to $U_P + U_I = 40 k_B T$. The backbones, as seen in Fig.~\ref{Fig5}A-F, clearly show that the branching density increases with decreasing $U_I$. $\langle Z \rangle$, (Fig.~\ref{Fig5}G) furthermore shows that the addition of a relatively small isotropic attraction ($U_I = 10 k_B T$) changes the initial aggregation kinetics to closely resemble that of the purely isotropic system, while still reaching the same long-time value of $\langle Z \rangle$ as in the purely directional case. By increasing $U_I$ further, the late-stage dynamics will also change, as particles can now start rearranging by escaping from their attractive wells defined by the strength of the directional bonds. The corresponding $\rho_L(\Lambda)$ curves (Fig.~\ref{Fig5}H) reflects the trends seen in Fig.~\ref{Fig5}A-F, and confirms that changing $U_I$ relative to $U_P$ allows for detailed control over the gel structure. The addition of a small isotropic attraction ($U_I = 10 k_B T$) causes a small \emph{increase} in branching, likely caused by the transition from partially reaction-limited to purely diffusion-limited aggregation. From there, increasing $U_I$ decreases the degree of branching due to the lateral rearrangement mechanism discussed earlier. Thus, using a combination between isotropic and directional interactions provides a route to preserving the diffusion-limited aggregation present in the isotropic model, while being able to reproduce the more branched networks as observed in experiment. Furthermore, tuning the relative magnitudes of isotropic and directional contributions allows fine-tuning of the network branching.

\section{Conclusions}
In this Article, we have described a computational model of colloidal gels formed through arrested spinodal decomposition based on particles with directional bonds (``patchy particles''). The simulation results were compared to experiments on gels formed from soft pNIPAm microgel particles through the use of a novel skeletonization algorithm that reconstructs a 3-dimensional image of the gel backbone from a collection of microscopy images, providing a versatile method for analyzing gel structures, including systems that do not offer single-particle resolvability in microscopy. Our results show that the use of attractive patches strongly promotes branching as compared to a purely isotropic model. This strong effect points towards the importance of inhibiting lateral particle rearrangements after initial aggregation due to surface roughness or entanglement between polymer strands on the touching particle surfaces. Overlaying isotropic and directional attractions furthermore allows for fine-tuning of the structure, which can be adjusted to resemble experimental gel structures. 

Unlike previous studies of the (equilibrium) phase behavior of patchy particles,~\cite{Sciortino2006,Zaccarelli2017} where the role of the directional bonds is to describe an underlying anisotropic interaction potential, the patchy potential here serves as a simple means of closing a \emph{kinetic} pathway towards gel coarsening present in the experimental system, even though the underlying interaction potential is approximately isotropic. In common with previous approaches based on three-body potentials,\cite{Kob2005} this induces an artificial short-range order dictated by the chosen patch geometry. One way to partially circumvent this drawback would be the use of \emph{randomly} distributed patches; an approach that was very recently investigated in Ref.~\citenum{Swan2019}. The use of such non-regular patch geometries however makes it difficult to prevent the formation of multiple bonds between patch pairs on the same two particles, and thus leading to a potentially ill-defined bond strength. The promising results presented here thus call for further investigations of the effects of different model approximations on the structure, coarsening dynamics, and rheology of model colloidal gels.

\section{Acknowledgments}
We gratefully acknowledge Minaspi Bantawa, Erika Eiser, Emanuela Del Gado, Stephen Hall, and Francesco Sciortino for valuable discussions. We gratefully acknowledge financial support from the European Research Council (ERC-339678-COMPASS) and the Swedish Research Council (grant numbers 2014-4037 and 2015-05449). Simulations were performed on resources provided by the Swedish National Infrastructure for Computing (SNIC) at LUNARC.

\section{Model and Methods}
\label{Sec-methods}

\subsection{Particle model}
The total pairwise interaction between microgel particles was described as the sum of a soft isotropic Hertzian repulsion $U_H$, given by
\begin{gather}
U_H(r) = 
\begin{cases}
\epsilon_H \left(1 - \frac{r}{\sigma}\right)^{\frac{5}{2}} & \text{if } r \leq \sigma \\
0 & \text{if } r > \sigma
\end{cases},
\label{eq-Hertz}
\end{gather}
and a rounded square-well like attraction $U_A$ describing short-range attractions, 
\begin{equation}
U_A(r) = \frac{\epsilon_A}{2}\tanh\left(\frac{r - w}{k}\right) - \frac{\epsilon_A}{2},
\label{eq-vdW}
\end{equation}
which was either isotropic or localized. In the above equations, $\sigma$ is the effective particle diameter, $\epsilon_H = 5 \times 10^3 k_BT$ sets the strength of the soft repulsion, with $k_B T$ the thermal energy, $r$ is the center-to-center distance between two particles (for $U_H$) or two patches (for $U_A$). In the attractive potential, $\epsilon_A$ describes the maximum attraction strength, while $w = \frac{\sigma}{20}$ and $k = \frac{\sigma}{200}$ specify the width and steepness of the attraction, respectively. The value of the repulsive strength $\epsilon_H$ was determined along the lines of previous work, after being adjusted to prohibit multiple-patch bonds,~\cite{Immink2019} while the attractive strength $\epsilon_A = 40k_BT$ was chosen large enough to ensure effectively irreversible particle bonds, in accordance with what we observe in experiments. The attractive, massless patches were distributed on the particle surface at positions corresponding to vertices of regular polyhedra, with each patch centered around point lying $0.4875\sigma$ outside the particle center, in order to prevent bonds between multiple patches. For the isotropic interaction case, the potential described by Eq.~\eqref{eq-vdW} instead acted over a spherical shell centered at the same distance from the particle center.

\subsection{Simulation details}
All simulations were performed using the open-source LAMMPS molecular dynamics package,~\cite{Plimpton1995} using overdamped Langevin dynamics for the translational and angular velocities $\textbf{v}_i$ and $\boldsymbol{\omega}_i$ of particle $i$:
\begin{eqnarray}
&\textbf{v}_i = \displaystyle \frac{\textbf{F}_i (\{ \textbf{r}_j \}) D_t}{k_B T} + \sqrt{2D_t}\boldsymbol{\eta}_t(t),\label{v_i}\\
&\boldsymbol{\omega}_i = \displaystyle \frac{\textbf{T}_i (\{ \textbf{r}_j \}) D_r}{k_B T} + \sqrt{2D_r}\boldsymbol{\eta}_r(t)\label{omega_i},
\end{eqnarray}
where the $\textbf{F}_i$ and $\textbf{T}_i$ is the force and torque on particle $i$ due to interparticle interactions, $D_t$ is the translational self-diffusion constant and $D_r= 3D_t/\sigma^2$ is the rotational diffusion, where we assume each particle to rotate as a solid sphere of diameter $\sigma$. $\boldsymbol{\eta}_t(t)$ and $\boldsymbol{\eta}_r(t)$ represent translational and rotational noise with unit variance, $\delta$-correlated in space and time. In practice, we achieved the dynamics in Eqs.~\eqref{v_i}--\eqref{omega_i} by solving the full (underdamped) Langevin equation, but with a damping time short enough to ensure that inertia is fully negligible, thus leading to effective overdamped dynamics. All simulations were run with $N = 10,000$ identical particles in a periodic box of linear dimensions $L=47.1\sigma$, yielding a volume fraction of $\phi = 0.05$ in order to mimic experiment. The initial configuration was one of randomly distributed particles, barring any unphysical particle overlap, thus resembling the experimental systems before aggregation due to particle collapse. All simulations were run four times with different initial configurations in order to obtain estimates of the variance in the different measured properties. Visualizations were obtained using the Open Visualization Tool OVITO.~\cite{Ovito}

In order to map the units of our simulations to experimental time and length scales, we rescaled all simulation and experimental results by $\sigma$ or the experimentally measured hydrodynamic diameter at 50~$^\circ$C (see below) and the Brownian time $\tau_B$, defined as
\begin{equation}
\tau_B \equiv \frac{\sigma^2}{D_t} = \frac{3\pi\mu\sigma^3}{k_B T}
\end{equation} 
where $\mu$ is the dynamic viscosity. The timestep was set to $2.4 \times 10^{-4} \tau_B$, corresponding to $1.2 \times 10^{-5}$ seconds at 50~$^\circ$C.

\subsection{Experimental procedures}
The particles were quenched instantaneously from 20~$^\circ$C to 50~$^\circ$C in $10^{-1}$M KCl, leading to the collapse of the individual particles followed by aggregation.~\cite{Zhu2011} The gels were allowed to rest for 10 minutes, after which they were imaged using CLSM. The pNIPAm particles used in this work were prepared in accordance with a previously described protocol.~\cite{Immink2019} Hydrodynamic diameters were determined from dynamic light scattering using a modulated 3D cross-correlation instrument (LS instruments) with a 660 nm diode-pumped laser, and were determined to 620 nm and 354 nm at 20~$^\circ$C and 50~$^\circ$C, respectively. It should be noted that the hydrodynamic radii were obtained at $10^{-3}$M KCl in order to prevent aggregation for temperatures above the volume phase transition temperature, leading to a slight overestimation of radii compared to fully screened conditions. The number density of the stock solution was determined at 20~$^\circ$C by counting the particles in at least 30 CLSM image stacks of fully crystalline samples. The number density together with the measured hydrodynamic radius was then used to calculate the volume fraction at 50~$^\circ$C.\cite{Royall2011a} A CLSM sample slide containing a $\phi = 0.05$ dispersion in $10^{-1}$M KCl was heated to 50~$^\circ$C by contact with a preheated metal block and imaged after 10 minutes using an inverted Leica TCS SP5 tandem scanner using a $100\times$ oil immersion objective, in an enclosure that allows for temperature control with a 0.2~$^\circ$C maximum variance using thermostatted air circulation.

\subsection{Skeletonization analysis}
Raw CLSM images for skeletonization were treated with Gaussian blurring using cubic kernels of size 5 voxel lengths (0.6$\sigma$) to remove speckled noise at gel surfaces, and subsequently binarized with thresholding according to Otsu's method.~\cite{Otsu1975} BD simulations were transformed into similar binarized image stacks from coordinate diagrams by projecting particle positions and sizes onto image slices. A morphological closing step was performed using a structural element of size 1 voxel, followed by removal of small unconnected structures ($<1$\% of largest cluster volume). These procedures involve Minkowski-based procedures adapted into MATLAB.\cite{Devaux2007} The skeletonization was then performed using previously developed algorithms~\cite{Chu1994,Fratzl2013} to create a pixel-thick backbone from binarized images. The skeleton was subsequently transformed into a set of nodes and links followed by a ``cleaning procedure'' which repeatedly (\emph{i}) removes branches connected to only one or two other nodes, excluding box edge nodes, and \emph{ii}) fuses nodes within a separation smaller than a particle diameter, until a structure emerges that cannot be cleaned further. For $\rho_L(\Lambda)$ calculations, links to the box edge nodes were not considered. 

\footnotesize{
\bibliography{bib_sim1}
}

\end{document}